\documentstyle[NATO,epsfig,graphics]{crckapb}
\vspace{4cm}
\begin{opening}
\title{Charge Distributions in Metallic Alloys: a Charge Excess Functional 
theory approach}
\author{Ezio Bruno}
\institute{\\ Dipartimento di Fisica and Unit{\`{a}} INFM,
Universit{\`{a}} di Messina,  Salita Sperone 31, 98166 Messina, Italy.
 E-mail: bruno@dsme01.unime.it. }
\end{opening}
\begin{document}
\section{INTRODUCTION}

In the last decade, the availability of large parallel computing resources 
and the development of order N algorithms~\cite{LSMS,LSGF} made feasible 
ab initio electronic structure calculations in extended metallic systems. 
A new, surprising, result in the theory of metallic alloys has 
been obtained by Faulkner, Wang and Stocks, who have analysed density functional 
theory calculations for unit cells containing hundred to thousand atoms and 
designed to simulate binary alloys with substitutional disorder. They 
discovered~\cite{FWS1,FWS2} that the net charge at each crystal site, $q_i$, 
is related to $V_i$, that part of electrostatic potential at the same site 
that is due to the interactions with all the other charges in the system, 
through a simple linear law  
\begin{equation}
\label{qvsv}
a_i q_i + V_i = k_i  
\end{equation}

For a specified configuration of the binary alloy A$_{c_A}$B$_{c_B}$, the 
coefficients $a_i$ and $k_i$ in Eq.~(\ref{qvsv}) take  the values $a_A$ 
and $k_A$ if the i-th site is occupied by a A atom or $a_B$ and $k_B$ 
otherwise. Moreover, the sets of coefficients extracted from different 
samples corresponding to the same mean concentration show up little differences. 
In the following, these linear relations shall be referred to as the $qV$ 
laws. The above new findings can be considered empirical in the sense that, 
although obtained from ab initio calculations, they have not yet been 
formally derived.

In spite of the simplicity of Eq.~(\ref{qvsv}), for each of the alloying 
species, the local charge excesses take {\it any} value within a certain interval. 
The corresponding distribution, even for the random alloy model, appears 
complex and cannot be reproduced, without a proliferation of adjustable 
parameters, in terms of the number of unlike neighbours of each 
site~\cite{Magri,Wolverton}. On the other hand, accurate calculations of 
the alloy total energies and phase equilibria must necessarily keep into 
account such a distribution. Recently, it has been shown that three 
coefficients of the $qV$ laws  for a binary alloy can be calculated within a 
single site theory, namely a Coherent Potential Approximation including 
local fields (CPA+LF)~\cite{CPALF}. More precisely, the coefficients 
$a_A$ and $a_B$ can be viewed as the responses of impurity sites occupied 
by A or B atoms to local external fields, while the third parameter can be 
viewed as the difference between the electronegativities of the A and B 
impurities embedded in the 'mean field alloy' defined by the 
alloy CPA Green's function.

\newpage
\setlength{\topmargin}{-1.3cm}
\setlength{\textheight}{25.5cm}
the local 
In the present paper, I shall demonstrate that the distribution of charges 
can be obtained from a variational principle, without any need of 
sophisticated electronic structure calculations for supercells. For this 
purpose I shall formulate a Ginzburg-Landau theory in which charge excesses, $q_i$, play the role of the order parameter field. 
Hereafter this phenomenological approach is referred to as the 'charge 
excess functional' (CEF) theory. As it will be seen below, the CEF is 
completely determined by only three concentration dependent, material 
specific, parameters. These parameters can be calculated by the CPA+LF 
theory or extracted from order N calculations. Given the atomic 
positions within a supercell, the CEF scheme determines the 
charge excesses at each site, and, hence, the electrostatic energy, 
with an excellent accuracy. Moreover, the above procedure, using a 
{\it single} set of parameters, can be applied to any ordered, partially 
ordered or disordered configuration corresponding to the same mean alloy 
concentration. Furthermore, CEF calculations require really modest 
computational efforts: 20 seconds CPU time on a 1 GHz Pentium III 
processor for a 1000 atoms sample. This is a particularly 
interesting feature as it opens new perspectives. In principle, one could 
take advantage from these performances and determine, in a parameter free 
theory, the equilibrium values of the short range order parameter for 
metallic alloys with an accuracy unprecedented for this kind of calculation. 
My group is currently developing a new computer simulation technique, based 
on the joint use of CEF and Metropolis' Monte Carlo that should allow the 
study of phase equilibria in metallic alloys.

The following of this paper is arranged as follows. In Section II, I shall 
present the CEF theory and its general solution for the site charge 
excesses. In Section III, the method will be applied to bcc 
Cu$_{0.50}$Zn$_{0.50}$ alloys and its accuracy will be tested through a 
comparison with order N Locally Self-consistent Multiple Scattering (LSMS) 
theory calculations~\cite{PCPA2001}. The comparison will also show that 
the CEF describes the distribution of 
charges in metallic alloys with a surprisingly good accuracy, when the 
material specific parameters are obtained from order N calculations, 
while fairly good results are obtained using the parameters obtained 
from the above generalisation of the CPA theory. The final Section IV is 
devoted to a thorough analysis of the CEF method and of its possible 
extensions and applications to the study of phase transitions and 
ordering phenomena in the metallic state.

\section{A CHARGE EXCESS FUNCTIONAL FORMALISM FOR CHARGE TRANSFERS IN 
METALLIC ALLOYS}
\subsection{The model}
The binary alloy A$_{c_A}$B$_{c_B}$, $c_A+c_B=1$, shall be studied by 
the means of supercells containing N 'atoms' with periodic boundary conditions. 
Each site of the cell can be occupied by a A or a B atom. If the chemical 
occupations are not considered, the lattice described by the 
sites within the supercell and their periodic replicas is a simple 
lattice, with one atom per unit cell. Below, it shall referred to as the 
'geometrical lattice'. In order to have a theory flexible enough to deal 
on equal footing both with ordered and disordered alloys, in principle, 
one should consider all the $\frac{N!}{(c_A N)! (c_B N)!}$ different 
'alloy configurations' that belong to the statistical ensemble specified by a 
given molar fraction, $c_A$. Each configuration is described by the set 
of 'occupation numbers', 
\begin{eqnarray}
\label{xi}
X_i^\alpha = \left\{ \begin{array}{ll} 
1 & \mbox{if the $i$-th site is occupied by a $\alpha$ atom} \\
0  & \mbox{otherwise} \end{array} \right.
\end{eqnarray} 
The arrays {\boldmath $X$}$^A$ and {\boldmath $X$}$^B$ describe 
completely an alloy configuration with a certain redundancy, since, 
for any $i$, it holds $X_i^A +X_i^B=1$.
Below, the convention is used that Latin indices, $i$, $j$, $\dots$, 
identify the sites in the supercell, and Greek indices, $\alpha$, $\beta$, 
$\dots$, the chemical species, A or B. Moreover, whenever their ranges are 
not indicated, the sums over the Latin indices run from $1$ to N, while 
the Greek indices take only the values A and B.

A volume $\omega_i$ is associated with each crystal site. In the 
following it will be assumed that all the atomic volumes sum up to the 
supercell volume. There is some arbitrariety in the way in which these 
volumes can be chosen: they could be built using the Wigner-Seitz 
construction (and possibly approximated by spheres, as 
in the case of the Atomic Sphere Approximation), or they could be 
non-overlapping muffin-tin spheres to which an appropriate fraction 
of the interstitial volume is added.

Each site in the supercell is occupied by a nucleus of charge $Z_i$, and, 
for each site, a charge excess can be defined as follows:
\begin{equation}
\label{qi}
q_i=\int_{\omega_i} d\vec{r} \rho(\vec{r}) - Z_i 
\end{equation}
where $\rho(\vec{r})$ is the electronic density. The above charge excesses 
satisfy a global electroneutrality condition
\begin{equation}
\label{electroneutrality}
\sum_i q_i =0
\end{equation}

Different models for ordered and disordered alloys will be discussed 
below. The {\it random alloy} model can be defined by saying the 
the occupations of different sites are not statistically correlated, i.e.
\begin{equation}
\label{ralloy}
\langle X_i^\alpha X_j^\beta \rangle=\langle X_i^\alpha \rangle 
\langle X_j^\beta \rangle
\end{equation}
or, equivalently, by assuming equal statistical weights for all the 
alloy configurations in a fixed concentration ensemble. This, 
evidently, corresponds to the $T \rightarrow \infty$ limit for the 
alloy site occupations. Real alloys, of course, should be studied at finite 
temperatures and, in order to describe {\it how much} they are ordered, 
it is customary to introduce the short range order 
parameters~\cite{WarrenCowley},
\begin{equation}
\label{sro}
p(\vec{r}_{ij})=\langle X_i^A X_j^B \rangle-\langle X_i^A \rangle 
\langle X_j^B \rangle
\end{equation}
Also a charge correlation function can be defined as
\begin{equation}
\label{chcorr}
g(\vec{r}_{ij})=\langle q_i q_j \rangle - \langle q_i \rangle \langle q_j \rangle
= \langle q_i q_j \rangle
\end{equation}
Of course, even for random alloys, the excess charges are correlated, i.e. 
$g(\vec{r}_{ij}) \ne 0$.

\subsection{The charge excess functional}
Within the muffin-tin or the atomic sphere approximation, the 
electrostatic energy of the system can be written as the sum of 
site-diagonal terms plus a Madelung term~\cite{Janak,FWS2}. I shall 
concentrate on the latter,
\begin{equation}
\label{emad}
E_M=\sum_{ij} M_{ij} q_i q_j=\frac{1}{2}\sum_i q_i V_i
\end{equation}
The Madelung matrix elements $M_{ij}$ in Eq.~(\ref{emad}) are 
defined~\cite{Ziman} as
\begin{equation}
\label{madmat}
M_{ij}=\sum_{\vec{R}} \frac{1}{|\vec{r}_{ij}+\vec{R}|}
\end{equation}
where $\vec{r}_{ij}$ are the translations from the 
$i$-th to the $j$-th site within the supercell and $\vec{R}$ are the 
superlattice translation vectors. 

Equation~(\ref{emad}) defines also the Madelung potential at the $i$-th 
site,
\begin{equation}
\label{vi}
V_i=2 \sum_j M_{ij} q_j 
\end{equation}
Everywhere in this paper atomic units are used in which $e^2=2$.

The starting point of the model is the {\it assumption} that linear laws hold 
and relate the charge excess at the i-th site $q_i$, and the Madelung potential 
at the same site $V_i$. As discussed above, this is an evidence from 
basically exact order N calculations, although also the single site CPA+LF 
model~\cite{CPALF} is able to provide a realistic estimate of the coefficients 
entering in the linear laws.  Therefore, I shall assume that, for  
some specified configuration, the following equations are satisfied,
\begin{equation}
\label{basic_eq}
a_i q_i + 2 \sum_j M_{ij} q_j= a_i b_i = k_i
\end{equation}
where $a_i$ and $b_i$, the coefficients of the linear laws, are 
assumed to depend only on the occupation of the $i$-th site in the 
configuration given 
and then to take the values $a_A$ and $b_A$ or $a_B$ and $b_B$ depending 
on the chemical occupation of the $i$-th site. Moreover, it is required 
that the global electroneutrality condition, Eq.~(\ref{electroneutrality}), 
must be satisfied. 

If all the material specific coefficients, $a_\alpha$ and $b_\alpha$, were 
specified by the mean alloy concentrations, one would have a set of N+1 
equations, Eqs.~(\ref{basic_eq}) and (\ref{electroneutrality}), and N 
unknown quantities to be determined, the $q_i$. In general, the determinant 
of this set of equations is not singular and, hence, the problem would be  
overdetermined. This is not in contrast with the results of order N 
calculations: in Refs.~\cite{FWS1,FWS2} it is found that all the four 
constants are determined {\it for a given configuration}, 
while different configurations corresponding to the same mean alloy 
concentration are characterised by slightly different sets of constants. 
Actually, in Ref.~\cite{FWS1}, the comparison between results for disordered 
and ordered alloy configurations shows that, for ordered alloys, the 
constants $a_\alpha$ have values very close to those found for random alloys 
at the same mean concentration, while larger discrepancies are found for the 
constants $b_\alpha$. 

As discussed in Ref.~\cite{CPALF}, an useful hint for solving the problem 
comes from the the CPA+LF model. This theory views the quantities 
$a_\alpha$ as the responses of the impurity sites, embedded in the CPA 
'mean' alloy, to a local field. Hence, in the CPA+LF model, the same quantities
depend {\it only} on the mean alloy concentrations. On the other hand, in 
the same theory, the zero-field charges, $b_A$ and $b_B$, are related one 
to the other through the CPA 'electronegativity' condition. These facts
suggest that, in different configurations corresponding 
to the same alloy concentration, the constants $b_\alpha$ are probably 
{\it renormalised} by the global constraint, Eq.~(\ref{electroneutrality}), 
while much smaller effects, if any, are expected for $a_\alpha$.

To make further progresses, consider the following functional of the site 
charge excesses, 
\begin{equation}
\label{functional}
\Omega([q],\mu)=\frac{1}{2}\sum_i a_i (q_i-b_i)^2 + \sum_{ij} M_{ij} 
q_i q_j - \mu \sum_i q_i
\end{equation}
where the Lagrange multiplier $\mu$ has been 
introduced to impose the global electroneutrality constraint.
By functional mimimization with respect to the order parameter field 
$\{q_i\}$, and to the multiplier $\mu$, the following set of Euler-Lagrange 
equations is obtained,
\begin{equation}
\label{el1}
a_i (q_i-b_i) + 2 \sum_j M_{ij} q_j= \mu 
\end{equation}
\begin{equation}
\label{el2}
\sum_i q_i=0 
\end{equation}
Equation~(\ref{el2}) evidently coincides with the electroneutrality
condition, Eq.~(\ref{electroneutrality}). On the other hand, 
Eq.~(\ref{el1}) reduces to Eq.~(\ref{basic_eq}) only when $\mu=0$. 
When $\mu \ne 0$, one can think that the renormalization of 
constants, $a_i b_i \rightarrow a_i b_i +\mu$ occurs in 
Eq.~(\ref{basic_eq}) in order to ensure the global electroneutrality 
constraint to be satisfied. If the problem of determining the site 
charge excesses is redefined as a minimum principle for the 
Ginzburg-Landau functional $\Omega$, 
the four constants $a_\alpha$, $b_\alpha$, obtained for {\it a 
given alloy configuration} can be used also for other configurations, 
since they will be properly renormalised: in other words, the 
information obtained from a specific configuration is {\it transferable} 
to other configurations belonging to the same fixed concentration ensemble. 

The minimum principle for  $\Omega$ leads to a scheme 
in which the constants $a_\alpha$, related with the response to the 
external potential, are not affected by the electroneutrality constraint. 
Now, since $\Omega$ has the dimension of an energy and contains the 
electrostatic energy, $\sum_{ij} M_{ij} q_i q_j$, one can think that the 
minimum of the functional 
\begin{equation}
\label{energy_functional}
E([q],\mu)=\frac{1}{2}\sum_i a_i (q_i-b_i)^2 + \sum_{ij} M_{ij} 
q_i q_j 
\end{equation}
corresponds the total electronic energy of the alloy configuration, except 
but an addictive constant. The 
quadratic terms in Eq.~(\ref{energy_functional}) can be considered as 
energy contributions associated with local charge rearrangements. 
Moreover, the quantity $\mu$, introduced simply as a Lagrange
multiplier, can be interpreted as a chemical potential ruling the 
charge transfers in metallic alloys. It is evidently related with the usual 
electronic chemical potential, $\mu_{el}$, that in the ensemble 
representable version of the density functional theory~\cite{erepr},
appropriate for random alloys, enters in the functional through the 
constraint,  
\begin{equation}
\label{erepr}
\int{d \vec{r} \; \rho(\vec{r})} =\sum_i Z_i 
\end{equation}

\subsection{Explicit determination of the charge transfers}
In the previous subsection, the functional $\Omega([q],\mu)$, hereafter 
referred to as the grand potential, has been introduced. Once minimised 
with respect to its variables, it provides the solution for the 
charge distribution and the chemical potential in the configuration considered, 
while the four constants entering in its definition, $a_\alpha$, $b_\alpha$, 
can be considered as the characteristic parameters 
for a specific alloy system at some specified concentration. In facts, the 
constants $b_\alpha$ can be evaluated for {\it any} alloy 
configuration corresponding to the given mean alloy concentration, while the 
arbitrariety introduced in this way is removed since they will enter in 
determining the charge distribution only through the combinations  $a_A 
b_A + \mu$ and $a_B b_B +\mu$, as 
discussed in the previous subsection. Below I shall elaborate 
the explicit solution of the problem. For this purpose, however, an 
appropriate formalism is necessary.

Below I shall use a tensor notation and denote 
the set of all the site charges, $q_i$, simply as {\boldmath $q$}. 
Analogously the sets of the Madelung potentials, $V_i$, and the site 
occupations, $X_i^\alpha$, shall be denoted as 
{\boldmath $V$} and {\boldmath $X$}$^\alpha$. Thus, e.g., 
{\boldmath $V$}=2 {\boldmath $M \cdot q$} should be read as $V_i=2 \sum_j 
M_{ij} q_j$. Moreover, I introduce the vector
\begin{equation}
\label{fvec}
\mbox{{\boldmath $f$}} = \sum_\alpha k_\alpha \mbox{{\boldmath $X$}}^\alpha 
\end{equation} 
and the tensor {\boldmath $\Gamma$} with matrix elements 
\begin{equation}
\label{gamma_tens}
\Gamma_{ij}= \sum_\alpha a_\alpha X_i^\alpha \; \delta_{ij}
\end{equation}
where $\delta_{ij}$ is the Kroenecker delta.
With this notation, Eqs.~(\ref{el1}) and (\ref{el2}) can be rewritten as
\begin{equation}
\label{el12}
(\mbox{\boldmath{$\Gamma$}}+ 2 \mathbf{M}) \cdot \mathbf{q}=  \mathbf{f} +\mu 
(\mbox{\boldmath{$X$}}^A +\mbox{\boldmath{$X$}}^B) 
\end{equation}
\begin{equation}
\label{el22}
(\mbox{\boldmath{$X$}}^A +\mbox{\boldmath{$X$}}^B) \cdot \mathbf{q}=0
\end{equation}

Then, the solution for the charge distribution can be written in terms of 
{\boldmath $\Lambda$} = ({\boldmath $\Gamma$}+2{\boldmath$M$})$^{-1}$, 
as follows
\begin{eqnarray}
\label{solq} 
\mbox{\boldmath{$q$}}  =  \mbox{\boldmath{$\Lambda$}} \cdot [
(\mbox{\boldmath{$f$}}+ \mu(\mbox{\boldmath{$X$}}^A
+\mbox{\boldmath{$X$}}^B)) ]
 =  (k_A +\mu) \mbox{\boldmath{$\Lambda$}} \cdot 
\mbox{\boldmath{$X$}}^A +
(k_B +\mu) \mbox{\boldmath{$\Lambda$}} \cdot \mbox{\boldmath{$X$}}^B
\end{eqnarray}
The chemical potential can be determined by multiplying Eq.~(\ref{solq}) on 
the left by ({\boldmath $X$}$^A$ +{\boldmath $X$}$^B$) and using 
Eq.~(\ref{el22}), e.g.,
\begin{equation}
\label{solmu}
(k_A +\mu) (\Lambda_{AA} +\Lambda_{BA})+
(k_B +\mu) (\Lambda_{AB} +\Lambda_{BB})=0
\end{equation}
where, the quantities
\begin{equation}
\label{lambdas}
\Lambda_{\alpha\beta} =\mbox{\boldmath{$X$}}^\alpha \cdot \mathbf{\Lambda} 
\cdot \mbox{\boldmath{$X$}}^\beta
\end{equation}
have been introduced.
Since the matrices $\Gamma_{ij}$ and $M_{ij}$ are real and symmetric, it 
follows that also $\Lambda_{ij}$ is real and symmetric and, hence,
\begin{equation}
\label{lambdasym}
\Lambda_{\alpha\beta} =\Lambda_{\beta \alpha} 
\end{equation}

By substitution of Eqs.~(\ref{solmu}) and (\ref{lambdasym}) in 
Eq.~(\ref{solq}), I find the final result of this section,
\begin{equation}
\label{solqfinal}
\mbox{\boldmath{$q$}}= (k_A - k_B) \; [ (1-y) 
\mbox{\boldmath{$\Lambda$}}
\cdot \mbox{\boldmath{$X$}}^A - y  \mbox{\boldmath{$\Lambda$}}
\cdot \mbox{\boldmath{$X$}}^B ]
\end{equation}
where
\begin{equation}
\label{x}
y=\frac{\Lambda_{AA}+\Lambda_{AB}}{\Lambda_{AA}+2 \Lambda_{AB}+\Lambda_{BB}}
\end{equation}
Eq.~(\ref{solqfinal}) clarifies that, in the general solution of the CEF for 
the charge distribution, $b_A$ and $b_B$ enter only via the difference 
$k_A - k_B=a_A b_A - a_B b_B$, while the dependence on the alloy 
configuration is conveyed by the quantity $y$.

The formulation of the charge excess functional, Eq.~(\ref{functional}), 
and the solution for the local charges, Eq.~(\ref{solqfinal}) are well 
defined both for ordered and disordered alloys. This has been possible 
because of the introduction of the chemical potential $\mu$: due to this, 
only {\it three} of the four constants that characterise the $qV$ linear 
laws enter in the final solution, Eq.~(\ref{solqfinal}). These 
three quantities together with $y$, determined by the actual alloy 
configuration, are equivalent to the original set of four constants.

In the next Section, the charge excess functional formalism will be applied 
to Cu$_{0.50}$Zn$_{0.50}$ alloys on a geometrical bcc lattice.
\begin{table}
\caption{Material specific parameters of the CEF, $a_{Cu}$, $a_{Zn}$ and 
$k_{Cu}-a_{Zn}$, for a bcc Cu$_{0.50}$Zn$_{0.50}$ alloy. The first set of 
parameters, indicated as LSMS, has been extracted from the $qV$ data 
obtained by the 'exact' LSMS calculations in Ref.[8] for a 1024 atoms 
supercell that simulate a random alloy. The parameters in the second set 
have been calculated using the CPA+LF model of Ref.[7]. All the quantities 
are in atomic units.}
\begin{center}
\begin{tabular}{ccc}

                            & LSMS     & CPA+LF \\
\hline
$a_{Cu}$                    & 1.84284   & 1.22787 \\ 
$a_{Zn}$                    & 1.82459   & 1.21890 \\
$k_{Cu}-k_{Zn}$             & 0.28957   & 0.14035 \\

\end{tabular}
\end{center}
\label{tabI}
\end{table} 

\section{CHARGE DISTRIBUTIONS IN BCC CuZn ALLOYS}
\subsection{Testing the CEF model}
As discussed above, Eqs.~(\ref{solqfinal}-\ref{x}) completely determine the 
distribution of charge excesses for a given alloy configuration. The three 
material specific parameters contained in the CEF can be 
extracted from order N as well as from CPA+LF calculations.
In this section I shall compare the charge excesses obtained by the CEF 
with order N Locally Self-consistent Multiple Scattering (LSMS) theory 
calculations~\cite{PCPA2001}. 
\begin{table}
\caption{CEF and CEF-CPA calculations (see the text for explanations) for 
the same bcc Cu$_{0.50}$Zn$_{0.50}$ sample as in Table I are compared with 
the 'exact' LSMS results of Ref.[8]. $\langle q \rangle_{Cu}$ and  $\langle 
V \rangle_{Cu}$ are, respectively, the mean values of the charges and the 
Madelung potentials at the Cu sites, $\sigma_{Cu}$ and $\sigma_{Zn}$ the 
standard deviations of the charge distributions for Cu and Zn, $E_{MAD}/N$ 
is the Madelung energy per atom and 'errors' stand for the mean square 
deviations between CEF (or CEF-CPA) charges and LSMS charges. All the 
quantities, unless otherwise stated, are in atomic units.}
\begin{center}
\begin{tabular}{cccc}
                            & CEF           & CEF-CPA & LSMS     \\
\hline
$\langle q \rangle_{Cu}$    & 0.099787      & 0.090649      & 0.099783   \\
$\langle V \rangle_{Cu}$ & -0.038197      & 0.039881      & 0.038188   \\
$\sigma_{Cu}$               & 0.02507       & 0.03082       & 0.02523    \\
$\sigma_{Zn}$               & 0.02801       & 0.03412       & 0.02814    \\
$E_{MAD}/N$   (mRy)         & -2.552        & -2.453        & -2.557     \\
'errors'                    & 2.7 10$^{-6}$ & 1.5 10$^{-4}$ &            \\
\end{tabular}
\end{center}
\label{tabII}
\end{table}

\begin{figure}
\centerline{\epsfig{figure=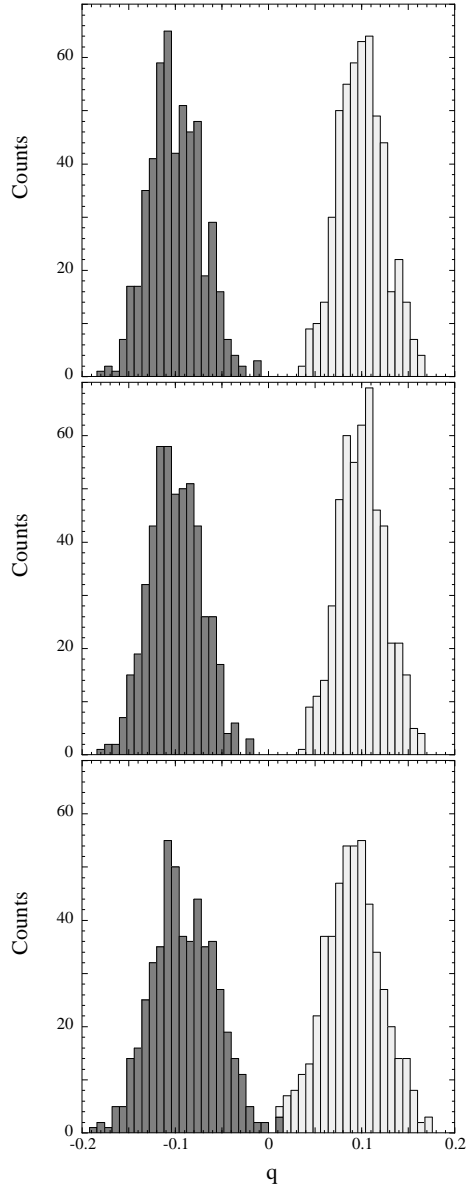,width=12cm}}
\caption{Cu (light histogram) and Zn (dark histogram) charge excesses 
distributions for the bcc Cu$_{0.50}$Zn$_{0.50}$ 1024 atoms sample of 
Table I. From top to bottom: LSMS results (Ref.[8]), CEF and CEF-CPA 
calculations. Atomic units are used.}
\label{histo}
\end{figure}
I have selected a specific configuration of a supercell containing 1024 atoms 
and designed to simulate a Cu$_{0.50}$Zn$_{0.50}$ random alloy on a bcc 
geometrical lattice for which LSMS calculations were available~\cite{PCPA2001}. 
Unless otherwise stated, the CEF calculations reported in the 
present paper have been performed on the same configuration. As reported in 
Table I, two distinct sets for the three CEF parameters (in this case 
$a_{Cu}$, $a_{Zn}$ and $k_{Cu} - k_{Zn}$) have been used. The parameters in 
the first set have been extracted from linear fits of the $qV$ data 
in Ref.~\cite{PCPA2001}, those in the second have been calculated by the 
CPA+LF model, as described in  
Ref.~\cite{CPALF}. Accordingly, two different 
sets of CEF calculations have been executed that shall be referred below 
to as: CEF, for the first set, or CEF-CPA, for the second.

\begin{figure}
\centerline{\epsfig{figure=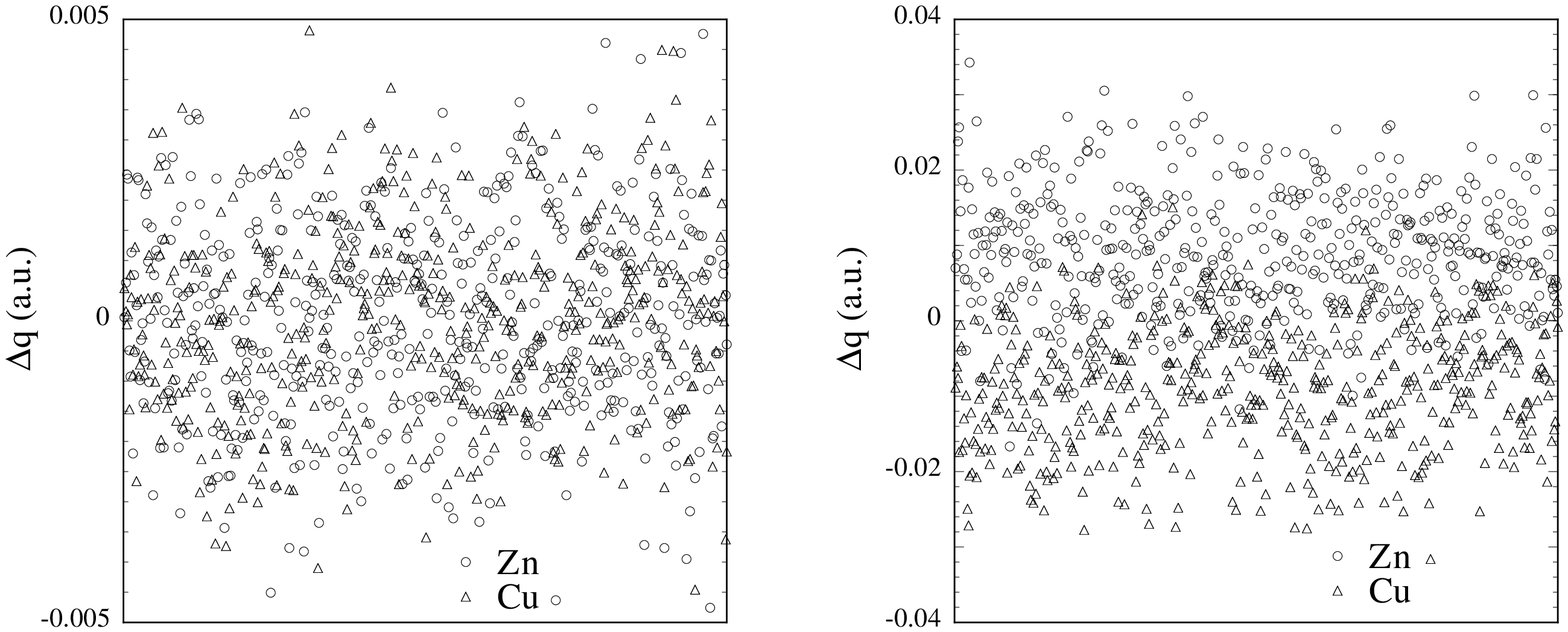,width=15cm}}
\caption{Left frame:  $\Delta q_i=q_i^{CEF}-q_i^{LSMS}$; right frame:
$\Delta q_i=q_i^{CEF-CPA}-q_i^{LSMS}$, where $q_i^{CEF}$, $q_i^{CEF-CPA}$ 
and $q_i^{LSMS}$ are, respectively, the site charge excesses at the $i$-th 
site as obtained by CEF, CEF-CPA or LSMS~(Ref.[8]) calculations for the 
bcc Cu$_{0.50}$Zn$_{0.50}$ 1024 atoms sample of Table I. In abscissa: site 
identifiers. Open circles: Zn sites; triangles: Cu sites. Atomic units are used.}
\label{fig2}
\end{figure}

The differences between LSMS and CEF calculations are really small,
as it is apparent from Table II: 5 parts over $10^5$ for the mean 
values of the charges and of the 
Madelung potentials, 2 parts over $10^4$ for the Madelung energies, less 
than 1 per cent for the widths of the charge distributions. The
distributions, reported in Fig.~(\ref{histo}), appear very similar. In 
order to have a more precise assessment of the accuracy of the CEF results, 
I have compared directly the charge excesses {\it at each lattice 
site}. In Fig.~(\ref{fig2}), 
the differences $\Delta q_i=q_i^{CEF}-q_i^{LSMS}$ are plotted.  
The absolute values of the $\Delta q_i$ are always smaller than 0.005 
electrons and no 
correlation is visible between the size of these 'errors' and the chemical 
occupation of the site. Interestingly, the mean square deviation between 
the two set of charges, reported in Table I, is of the order of $10^{-6}$, 
i.e. its size is comparable with the numerical errors in LSMS calculations. 
The main source of the tiny differences found is that all the CEF charges,
{\it by construction}, lie on the two straight lines corresponding to 
the $qV$ laws for each of the alloying species, while the same laws hold 
only approximately for LSMS calculations. Tests about the 
{\it transferability} of the CEF parameters extracted from one sample to the 
other samples are currently being performed. Preliminary results already 
available~\cite{CEF} suggest that, when using parameters extracted from one 
sample, the CEF is able to reproduce the charge distribution in other samples 
maintaining the above mentioned accuracy,~~even when CEF parameters 
extracted~~from random samples are used in ordered,
partially ordered or segregated samples and vice versa. This 
{\it transferability} of the CEF parameters is a very remarkable result:  
evidently the renormalization of the constants mentioned in the previous 
section is able to deal with very different samples and, more important, 
this implies that the CEF theory is generally applicable to 
metallic alloys, no matters whether they are ordered, 
disordered or segregated. Furthermore, the accuracy obtained for the Madelung 
energy demonstrates that the theory is able to describe very carefully the 
electrostatic contributions to the energetics of ordering phenomena.

The application of the CEF theory using parameters from CPA+LF 
calculations has a particular interest since the CPA+LF model is not based on 
a specific alloy configuration and, therefore, does not require expensive 
calculations on supercells. In this case, as it can be seen in Table I 
and Figs.~(\ref{histo}-\ref{fig2}), the agreement with LSMS calculation is 
still fairly good: the CEF-CPA understimates the mean charges about 10 
per cent and overestimates the widths of the charge distributions about 
25 per cent. These errors somehow compensate 
giving a Madelung energy correct within 4 per cent. The comparison with 
LSMS for the charges at each sites, as displayed in Fig.~(\ref{fig2}), 
shows up 
small systematic errors with different signs on Cu and Zn sites. The 
histograms of the charge distributions present a small overlap around 
$q=0$, as it is visible in Fig.~(\ref{histo}). Also in this case, as 
above, preliminary 
tests~\cite{CEF} shows that the parameters obtained from the CPA+LF 
theory are {\it transferable}, in the 
sense that the size of the discrepancies between CEF-CPA and LSMS results 
appear independent on the amount of short range order in the alloy 
configurations considered. To maintain the same performances, even in the 
cases of ordered 
or segregated samples, is a very remarkable success for 
a theory, the CPA, originally proposed for random alloys. To my knowledge, 
this is the first time that a single site theory, free of adjustable 
parameters, is able to reproduce the charge distribution in metallic alloys. 
Better results have been obtained by the Polymorphous CPA 
(PCPA)~\cite{Ujfalussy} that, 
although based on the CPA  theory, at similarity of 'exact' LSMS calculations, 
uses supercells and, hence, many different site potentials. 

\begin{figure}
\centerline{\epsfig{figure=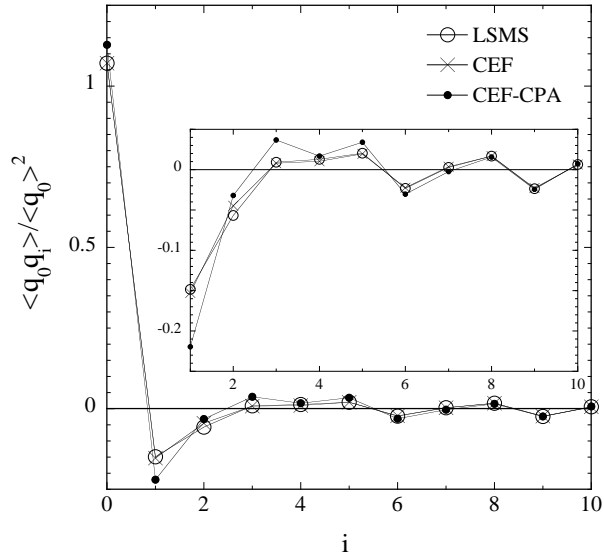,width=10cm}}
\caption{Normalised charge correlation function $\langle q_0 q_j \rangle/ 
\langle q \rangle_{Cu}^2$ for the 1024 atoms bcc Cu$_{0.50}$Zn$_{0.50}$ random alloy 
sample (see Table I). Open circles: LSMS calculations~(Ref.[8]), crosses: 
CEF calculations, filled circles: CEF-CPA calculations. In the inset,  
a detail of the same curve is plotted. The shell 
identifyer, $i$, is reported in abscissa. Lines joint the points.}
\label{fig3}
\end{figure}
An important quantity, that is particularly relevant for its role in the 
energetics of metallic alloys~\cite{Wolverton}, is the charge correlation 
function $g(\vec{r}_{ij})$. In Fig.~(\ref{fig3}), I plot $g(\vec{r}_{ij})$, 
as obtained from LSMS, CEF and CEF-CPA calculations, again for 
the 1024 atoms Cu$_{0.50}$Zn$_{0.50}$ random alloy sample of 
Ref.~\cite{PCPA2001}. As it is apparent, the agreement between LSMS and 
CEF calculations is excellent, and very good also for the CEF-CPA. The test 
is particularly interesting since non correlated charge models would give 
$g(\vec{r}_{ij})=0$ for $r_{ij}>0$. It could be therefore quite 
surprising to observe that, at least in the case at hand, the correlations are 
slightly {\it overestimated} by the CEF-CPA model.

\subsection{Charge excesses versus local environments}
The importance of local environments in determining the charge transfers in 
metallic alloys has been highlighted for the first time by Magri et 
al.~\cite{Magri}. Their model simply assumes the charge excesses to be 
proportional to the number of {\it unlike} nearest neighbours. When the 
development of order N calculation allowed  a deeper investigation of 
the problem~\cite{FWS1,FWS2}, it was readily clear that such a simple 
model  was not able to describe the details of the charge 
distributions. Later on, however, Wolverton et al.~\cite{Wolverton} 
generalised Magri's model by introducing additional terms proportional to 
the number of neighbour in outer shells and achieved appreciable improvements 
especially for fcc lattices. 

\begin{figure}
\centerline{\epsfig{figure=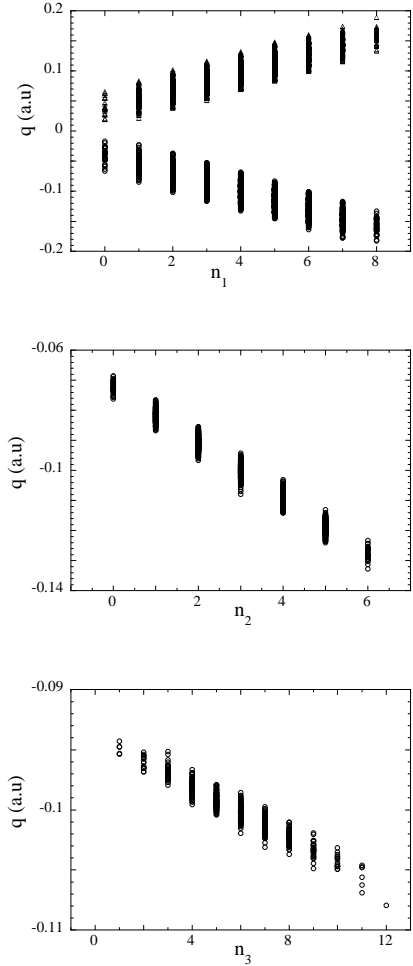,width=10cm}}
\caption{Top frame: charge excesses, $q$, vs. the number of unlike nearest 
neighbours, $n_1$, for bcc Cu$_{0.50}$Zn$_{0.50}$ random alloys. 
Middle frame: $q$ vs. the number of unlike neighbours in the {\it 
second shell}, $n_2$, {\it only} for the Zn atoms that have 4 unlike 
neighbours in the first shell. Lower frame: $q$ vs. the number of 
unlike neighbours in the {\it third shell}, $n_3$, {\it only} for 
the Zn atoms that have 4 and 3 unlike neighbours in the first two shells.
Open circles and triangles identify, respectively, Zn and Cu 
sites. The data plotted have been obtained from CEF calculations on 40
random alloy samples each containing 432 atoms on a geometrical bcc 
lattice. The CEF parameters used are listed in Table I. Atomic units are 
used.}
\label{alex}
\end{figure}

The computational flexibility of the CEF method and the fact that it seems 
able to reproduce almost perfectly LSMS results allow to check the basic 
assumptions of the class of theories to which the models of 
Refs.~\cite{Magri,Wolverton} belong. For this purpose I have 
evaluated the charge distributions in 40 Cu$_{0.50}$Zn$_{0.50}$ bcc 
random alloy samples, each containing 432 atoms. The CEF parameters have 
been extracted from the LSMS calculations of Ref.~\cite{PCPA2001} (see 
Table I). The data obtained are analysed in Fig.~(\ref{alex}).
In the top 
frame the individual charge excesses are plotted vs. the number of nearest 
neighbours, $n_1$. The existence of correlation between the charge 
excesses, $q$, both with the site occupation and $n_1$, is evident. However, 
as it is apparent, the excesses of charge for atoms 
of the same chemical species and the same number of unlike nearest 
neighbours can take any value in intervals whose typical widths is about 
0.05 electrons, moreover intervals corresponding to different values 
for $n_1$ present appreciable overlaps. The same observations have already 
been made in Ref.~\cite{FWS1}, the main difference is that I have considered a 
much larger number of configurations and used sample all corresponding to the 
same stoichiometry. 
The conclusion, however remains the same: $n_1$ is not 
sufficient to characterise the distribution of $q$. In order to check how 
much the consideration of the number of neighbours in the second or in the 
third shell, $n_2$ and $n_3$, can improve, I have selected all the Zn atoms 
with $n_1=4$ and plotted their charges vs. $n_2$ (Fig.~(\ref{alex}), middle 
frame) and all the Zn site with $n_1=4$ and with $n_2=3$, the corresponding 
charges are plotted in the lower frame of the same Fig.~(\ref{alex}). 
Although the qualitative picture is not changed, it is clear that, if the 
occupation of the neighbours in the first three shells is known, the 
uncertainty on the charge is reduced about one order of magnitude with 
respect to what can be obtained by considering $n_1$ only. In any case,  
trying to improve Magri's model by including more and more shells and more 
and more adjustable parameters, in my view, appears misleading in that it 
obscures the simple fact that a single number, the value of the Madelung 
field, is able to reduce the uncertainty to an amount comparable 
with numerical errors in order N calculations. 

\section{DISCUSSION}
I like to conclude this paper by making, in a quite sparse order, 
some comments about several interesting aspects of the CEF model and 
discussing about possible future applications of the theory.

{\bf $i$)~A Coarse graining over the electronic degrees of freedom.}~~The 
CEF operates a {\it coarse graining} over the electronic degrees 
of freedom, that are reduced to {\it one for each atom}, the local excess of charge,
{\it without any appreciable loss of accuracy} for the total energy. 
This is a consequence of the fact that, within theories like the 
CPA+LF~\cite{CPALF} or the PCPA~\cite{Ujfalussy} any site diagonal property is a 
{\it unique function} of the Madelung potential, $V_i$, and the nuclear charge at 
the same site, $Z_i$. As noticed in~\cite{CPALF} this uniqueness is due to the 
mathematical simplicity of the CPA projectors and, therefore, probably does 
not hold for more exact approaches, where some residual dependence on the 
site nearest neighbours environment is expected for. Nevertheless, the fact 
that the CPA theory accurately accounts for the spectral properties of 
metallic alloys~\cite{Abrikosov_cpa} and the quantitative agreement with LSMS 
calculations found in Refs.~\cite{CPALF,Ujfalussy} suggest that the errors 
introduced by neglecting the nearest neighbours influence are probably 
not much larger than numerical errors in order N electronic structure 
calculations. The previous sentences require some clarification: when 
referring to the nearest neighbours environment, I mean the effects of the 
environment not already conveyed by $V_i$. The site Madelung potential, in 
fact, already contains much information about the occupations of near and 
far sites, each weighted as appropriate. Although the context was 
very different, I like to recall that a coarse graining over quantum 
degrees of freedom has precedents in the concepts of chemical valence and, 
more quantitatively, in that of electronegativity.  

{\bf $ii$)~The CEF and the local environments.}~~Previous attempts to build 
theories dealing with charge transfers in 
metallic alloys, as, for instance the model of Magri et al.~\cite{Magri} or 
the charge-correlated CPA of Johnson and Pinski~\cite{more_scr}, have been 
focused on the number of unlike neighbours of each site. Subsequent 
extensions~\cite{Wolverton} included consideration for the occupations of 
outer shells. The $qV$ linear relations suggest that the convergence 
of such schemes in the number of shells is slow, being basically 
related to the $r^{-1}$ decay of the Coulombian interaction. The CEF 
model is more effective in that it accounts for these long ranged 
interactions. This notwithstanding, the models of 
Refs.~\cite{Magri,Wolverton} 
suggest routes to possible future refinements of the CEF theory: 
improvements could be obtained, for instance, by including local 
fields in the charge correlated CPA model of Ref.~\cite{more_scr}.

{\bf $iii$)~Computational performances of the CEF theory.}~~Modern 
ab initio order N calculation require a number of 
operations directly proportional to the number of atoms in the 
supercell, N, unfortunately with huge praefactors. To fix the ideas, 
consider the case of LSMS calculations: the number of operations 
required is given by
\begin{equation}
n_{LSMS} \propto n_L^3 \; n_{LIZ}^3 \; n_E \; n_{it} \; N 
\approx 6 \cdot 10^{9} N\end{equation}
where $n_L=(l_{max}+1)^2$ is the size of the single site scattering 
matrices, $n_{LIZ}$ the number of atoms in the local interaction 
zone, $n_E$ and $n_{it}$ the number of points in the energy mesh 
and the number of iterations that are necessary to solve the 
Kohn-Sham equation. The above estimate for the praefactor is quite 
optimistic and correspond to assuming $l_{max}=3$, 
$n_{LIZ}=24$, $n_E=n_{it}=10$.
The CEF requires $n_{CEF}=N^3$ operations when using conventional 
linear algebra algorithms. Accordingly, for a typical size of the 
supercell, $N=1000$, the CEF is 3 to 4 orders of magnitude faster than 
LSMS. Of course, order N matrix inversion algorithms can be used for 
the CEF also, this would give $n_{CEF}= n_{LIZ}^3 N$, i.e. 5 orders 
of magnitude faster than LSMS, regardless of the size of the supercell.

{\bf $iv$)~A CEF-Monte Carlo mixed scheme.}~~The very remarkable speed 
up in electronic structure calculations that can be 
obtained using the CEF has a qualitative relevance because it opens 
unexplored possibilities. 
The minimum value of the CEF functional for a given alloy configuration, 
$X$, can be viewed as the total electronic energy corresponding to that 
configuration. Therefore, the same minimum value can be regarded as 
a {\it functional of the alloy configuration}, say:
\begin{equation}
Min_{\{q\}} \Omega([q]; X, c)=E_{el} (X; c)
\end{equation}
If lattice vibrations and deformations are not considered, $X$ is
completely equivalent to the whole set of the atomic positions. 
If the validity of the Born-Oppenheimer approximation 
and of a classical approximation for the atomic degrees of freedom are 
assumed, then $E_{el} (X; c)$ can be regarded as a classical 
Hamiltonian for the alloy in study. Probably the functional dependence of 
$E_{el} (X; c)$ on the atomic degrees of freedom, X, is too much 
complicated for exact, even though approximate, statistical studies. 
My group is currently developing a mixed CEF-Monte Carlo scheme in which 
a Metropolis Monte Carlo algorithm is used to obtain ensemble averages for 
the classical Hamiltonian $E_{el} (X; c)$. The goal here is being able to 
determine {\it ab initio}, within {\it a non perturbative} method, the 
thermodynamics, the phase equilibria and the atomic correlation functions
for metallic alloys. At the same time, taking advantage of the uniqueness 
of the site properties within CPA based approaches, such a scheme should allow for 
a careful determination of the electronic properties along the lines of the 
LSMS-CPA of Ref.~\cite{sam98}. 

{\bf $v$)~Improving the CEF-CPA.}~~In Section III, the distributions of the site charge excesses 
in a random alloy system have been studied. The validity of the $qV$ laws 
implies that also the values of the Madelung field, $V$, at different sites 
can be described by two distributions 
$d_\alpha(V)$. With respect to these, a random alloy system 
can be viewed as a {\it charge glass}. In fact, the consequences of the 
$q$ and $V$ {\it polydispersivity} on the energetics of random 
alloys are similar to those of the polydispersivity of the bond lengths 
in ordinary glasses. It is easy to see that these distribution satisfy 
the following sum rules: 
\begin{eqnarray} \nonumber
& & \int_{-\infty}^{\infty} { d V \; d_\alpha (V)} =1 \\
& & \sum_\alpha c_\alpha \int_{-\infty}^{\infty} { d V \; V \; d_\alpha (V)}=0
\label{dphi}
\end{eqnarray}
On the other hand, the standard CPA theory is based on the implicit 
assumption that 
\begin{equation}
d_A(V)=d_B(V)=\delta(V)
\label{dphicpa}
\end{equation}
i.e., the CPA considers random alloys are as  {\it charge monodisperse} 
systems. Therefore, the CEF-CPA scheme presents the 
inconsistency that, while the parameters entering in the CEF are 
calculated by assuming the distribution in Eq.~(\ref{dphicpa}), the output  
distributions are typical of charge glasses. 
As it is well known, appreciable 
improvements over the standard CPA theory can be achieved by the 
SIM-CPA~\cite{Abrikosov} or the screened CPA~\cite{more_scr} models. Both 
theories are based on the prescription $d_A(V)=\delta(V-V_A)$,  
$d_B(V)=\delta(V-V_B)$, where the $V_\alpha$ are chosen in order to mimic the 
{\it mean} effect of the charge correlations,
in such a way that the sum rules are obeyed. Although these are 
still {\it monodisperse} theories, displacing the centre of mass 
of the distributions allows for substantial improvements. 
The best CPA-based model to date available, the 
PCPA of Ujfalussy et al.~\cite{Ujfalussy} is a truly {\it polydisperse} 
theory in which the $V$ distributions 
are defined self-consistently by the supercell used. As a theory based on specific 
supercells, however, the PCPA cannot (at least, without much labour) make 
predictions on the atomic correlations. 
We are currently developing an alternative approach that could maintain the advantages of  
the CEF-Monte Carlo scheme without paying the price of having non consistent 
charge distributions. The idea is simple: an approximation very similar to 
the PCPA theory  can evaluate the 
polymorphous model defined by the $V$ distributions obtained as an output of 
the CEF-Monte Carlo, rather those defined by a specified supercell. In this way a new 
set of improved coefficients for the CEF can be obtained. Thus, by 
iterating the above modified PCPA and the CEF-Monte Carlo, until convergence 
is obtained for the $V$ distributions, one would obtain a completely ab initio 
non perturbative quantum theory of metallic alloys able to evaluate, 
at the same time, electronic and atomistic properties. 

\begin{acknowledgements}
I wish to thank Sam Faulkner and Yang Wang that made available in digital 
form the data of Ref.~\cite{PCPA2001}. I also aknowledge many interesting 
discussions with Sandro Giuliano, Antonio Milici and Leon Zingales.
\end{acknowledgements}

\end{document}